# Ion-induced transient potential fluctuations facilitate pore formation and cation transport through lipid membranes

D. Roesel[a,1], M. Eremchev[a,1], C. S. Poojari[b], J. S. Hub[b,2], and S. Roke[a,c,d,2]

[a]Laboratory for Fundamental BioPhotonics (LBP), Institute of Bioengineering (IBI), School of Engineering (STI), École Polytechnique Fédérale de Lausanne (EPFL), CH-1015 Lausanne, Switzerland; [b]Theoretical Physics and Center for Biophysics, Saarland University, 66123 Saarbrücken, Germany; [c]Institute of Materials Science and Engineering (IMX), School of Engineering (STI), École Polytechnique Fédérale de Lausanne (EPFL), CH-1015 Lausanne, Switzerland; and [d]Lausanne Centre for Ultrafast Science, École Polytechnique Fédérale de Lausanne (EPFL), CH-1015 Lausanne, Switzerland

[1] D.R. and M.E. contributed equally to this work.

[2]To whom correspondence may be addressed. Email: jochen.hub@uni-saarland.de or sylvie.roke@epfl.ch.

**Abstract**

Unassisted ion transport through lipid membranes plays a crucial role in many cell functions without which life would not be possible, yet the precise mechanism behind the process remains unknown due to its molecular complexity. Here, we demonstrate a direct link between membrane potential fluctuations and divalent ion transport. High-throughput wide-field second harmonic (SH) microscopy shows that membrane potential fluctuations are universally found in lipid bilayer systems. Molecular dynamics simulations reveal that such variations in membrane potential reduce the free energy cost of transient pore formation and increase the ion flux across an open pore. These transient pores can act as conduits for ion transport, which we SH image for a series of divalent cations ($Cu^{2+}$, $Ca^{2+}$, $Ba^{2+}$, $Mg^{2+}$) passing through GUV membranes. Combining the experimental and computational results, we show that permeation through pores formed via an ion-induced electrostatic field is a viable mechanism for unassisted ion transport.

**Introduction**

Ion transport through lipid bilayer membranes is necessary for numerous vital functions, essential for life. Ion transport occurs either actively, using specific channels and pumps, or passively, by virtue of physical processes that render membranes permeable for ions. Ion transport has been studied in many different ways, such as fluorescence microscopy[1,2], isotopic flux measurements[3,4], and electrical measurements[5,6]. Active ion transport facilitates highly specific functions, while passive ion transport is necessary to maintain the general physiological equilibrated state of the cell. However, because of the molecular complexity, the precise molecular mechanism for passive ion transport remains unknown. Several mean-field mechanisms have been proposed such as thermal pore formation, solubility-diffusion, lipid flip-flop, and ion-induced pore formation, explained in detail in Supplementary Section S1 and Fig. S1-S3, but until now it is not clear what is the dominating mechanism. Likewise, it has been pointed out that lipid membranes harbor a unique structural heterogeneity, which is ascribed to having well-defined – but mostly unknown – functions. Therefore, it is of great interest to achieve an understanding of ion transport that goes beyond the mean-field level, using experimental methods that are sensitive to molecular interfacial structure in a space and time-resolved fashion that is relevant for the ion transport process.

High throughput second harmonic imaging[7–9] has been introduced as a label-free method using the detection of water at buried interfaces in aqueous solution, such as interfacial water inside a glass micro-capillary[10], or interfacial water in contact with lipid membranes[11–13]. In a non-resonant SH imaging experiment a pulsed femtosecond near-infrared laser beam interacts with an aqueous interface. Coherent SH photons are emitted from all non-centrosymmetric molecules that are non-centrosymmetrically distributed. Since interfacial water is non-centrosymmetrically distributed, while isotropic bulk liquid is not, high throughput SH imaging has an exquisite interfacial sensitivity (Fig. 1A). The interfacial water outnumbers lipids with a ratio of >1:100[14,15], and due to the non-resonant nature of the process, molecules of both constituents respond with electromagnetic SH fields that have equal magnitudes. Because the SH intensity scales quadratically with the emitted electromagnetic field, which is proportional to the number of anisotropic source molecules, the SH images generally report on membrane hydration, detecting the net orientational distribution of water molecules along the surface normal. The response of this oriented water has been successfully used to quantify physio-chemical interfacial properties, such as the electrostatic potential ($\Phi_0$). Using high-throughput SH imaging we recently observed membrane potential fluctuations on free-standing planar extended lipid membranes embedded in aqueous solution[11].

For biotechnological or medical applications, external electric fields are widely used to induce pores in membranes to enable the uptake of genes, drugs, or other cargos by biological cells[16,17]. This process, called electroporation, has been used and studied in great detail by experimental, theoretical, and computational methods[18–20]. In molecular dynamics (MD) simulations, since pores have been induced under non-equilibrium conditions by the application of excessive transmembrane potentials, it has been difficult to compute the effect of electric fields on the free energies of pore formation, and the source of these transmembrane potentials has not been included in the model. Therefore, it has remained unclear whether transmembrane potentials formed *intrinsically* in the membrane, i.e., through the binding of ions to the lipid head groups, could play any role at all in terms of passive ion transport.

Here, we show that membrane potential fluctuations are universally found in free standing membranes as well as in giant unilamellar vesicles. Non-resonant SH imaging data shows that they exist on charge-asymmetric membranes, induced either by lipid asymmetry or by ion asymmetry, as well as on symmetric membranes, induced by the spatiotemporal structural fluctuations. Exploring such variations using molecular dynamics simulations, we find that transient membrane potential fluctuations (on the order of several hundreds of millivolts) reduce the free energy for the formation of transient pores. These transient pores act as conduits for (di)valent ion transport, which we observe for a series of divalent cations ($Cu^{2+}$, $Ca^{2+}$, $Ba^{2+}$, $Mg^{2+}$) passing through GUV membranes composed of unsaturated lipids. The average translocation rate correlates with the average transient membrane potentials and agrees well with the permeabilities found from MD simulations. Therefore, we propose permeation through pores formed via an ion-induced electrostatic field as a viable mechanism for unassisted ion transport.

**Results and Discussion**
**Membrane potential fluctuations and transient pore formation.**

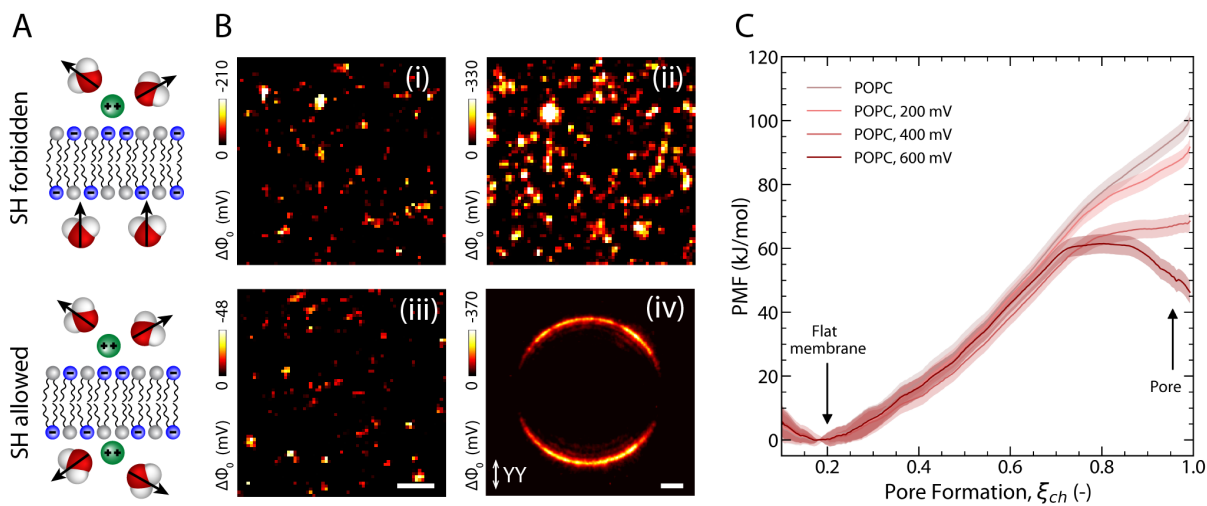

**Figure 1: Membrane potential fluctuations.** (A) Top: Schematic of divalent cations binding to anionic lipids which disturbs the water orientation around one leaflet of the membrane, thus making the interface SH active. Bottom: divalent cation translocation disturbs the water orientation on both sides making the interface SH inactive. (B) Membrane potential $\Delta\Phi_0$ snapshots obtained from FLMs and GUVs with different sources of asymmetry: (i) asymmetric lipid composition (FLM, DPhPC:DPhPA 70:30 / DPhPC in 150 μM KCl), (ii) asymmetric solutions (FLM, DPhPC:DPhPA 70:30 in 50 μM CaCl$_2$ / 150 μM KCl), (iii) symmetric lipid composition and solutions (FLM, DPhPC:DPhPA 70:30 in 50 μM CaCl$_2$), and (iv) asymmetric solutions (GUV, DPhPC:DPhPA 1:1 in 5 mM CaCl$_2$). (C) Potentials of mean force (PMFs) of pore formation over POPC membranes at transmembrane potentials between 0 and 600 mV, computed with the ECC force field (see legend).

Figure 1B shows membrane potential ($\Delta\Phi_0$) snapshots obtained from free standing lipid membranes (FLMs) composed of (i) a mixture of DPhPC:DPhPA 70:30 top leaflet and DPhPC bottom leaflet surrounded by an aqueous 150 μM solution of KCl, (ii) a symmetric DPhPC:DPhPA 70:30 membrane with the bottom leaflet in contact with an aqueous solution of 50 μM CaCl$_2$ and the top leaflet in contact with an aqueous solution of 150 μM KCl, (iii) a symmetric DPhPC:DPhPA 70:30 membrane embedded in a 50 μM solution of CaCl$_2$, (iv) shows a membrane potential snapshot of a GUV with a solution of 45 mM sucrose on the inside and 5 mM CaCl$_2$ with 30 mM glucose on the outside, see Materials and Methods for more details on sample preparation and Supplementary Section S2 for membrane potential conversion. Note that the phospholipids were chosen to have branched alkyl chains, which makes them resistant to ion transport. All these images display intensity and thus membrane potential fluctuations with magnitudes on the order of several hundred millivolts for each micrometric domain. The temporal and spatial fluctuations observed on the GUV are presented in more detail in the Supplementary Section S3 (Fig. S4). While the presence of potential fluctuations in a planar extended lipid membrane might still be due to the specific properties of the membrane model system, the presence of such fluctuations under different conditions and in vastly different model systems suggests that the membrane potential and thus free energy landscape of charged lipid membranes is a general dynamic feature of phospholipid membranes.

We expect that the source of these fluctuations is found in the charge condensation layer that is composed of the hydrated lipid headgroup region. This region consists of the charged head groups themselves as well as their counterions, and also includes ionic species from the bulk aqueous phase. For a typical 70:30 mole fraction membrane with a transmembrane potential difference of ~200 mV the degree of ionisation is ~1 %[11] and the local concentration of ions is 1–3 M[14], well above the Kirkwood transition[21,22] of ionic solutions (60 mM). Above this concentration the distribution of ions is no longer statistical. X-ray diffraction experiments dating back to the 1930's[23,24] have shown that at such high concentrations ions may even be distributed in quasi-periodic lattices spaced by more dilute regions of ions[25]. More recent dynamic light scattering studies report on larger sub-micron sized domains[26], on the length scale of 0.1–0.5 microns. Given that high concentrations of ions are present in the charge condensed layer that is the hydrated headgroup region of a lipid bilayer, we expect that a similar clustering occurs and that these transient clusters are responsible for the observed membrane potential fluctuations. Since this ion clustering is a general phenomenon, the free energy variations should be universally present on charged lipid membranes in aqueous solution, in agreement with our data.

We therefore investigated the effect of transmembrane potentials on membrane permeability with all-atom MD simulations and free energy calculations (See Materials and Methods for details). We computed the potential of mean force (PMF, sometimes referred to as free energy profile) for the formation of an aqueous defect over membranes at transmembrane voltages between 0 and –600 mV (Fig. 1C, Methods). Membranes of POPC were simulated to enable the use of force fields with refined lipid-ion interactions[27–29]. The PMFs were computed along the chain reaction coordinate $\xi_{ch}$, which quantifies the degree of connectivity along a polar transmembrane defect, where $\xi_{ch}\approx0.2$ and $\xi_{ch}\approx0.9$ correspond to the states with a flat membrane and with a continuous transmembrane defect[30]. In contrast to previous MD studies[18,19], by using umbrella sampling along $\xi_{ch}$, we could simulate the pore opening pathway under equilibrium conditions and, thereby, obtain accurate free energies of pore formation. The PMFs reveal that, while pores are highly unfavorable in the absence of a potential, potential magnitudes up to 600 mV greatly stabilize the pores by up to 40 kJ/mol. In addition, the potentials may render the pores metastable (long-living on the molecular time scale), as evident from the emergence of a local free energy minimum at $\xi_{ch}\approx1$ (Fig. 1C, dark red).

Observing such a massive reduction in the free energy of transient pores, we next investigate if such pores are permitting divalent cation transport. To do so, we first examined the effect of a hydrophobic core on pore formation (Supplementary Section S4, Fig. S5), comparing DPhPC and DOPC membranes in MD simulations. Fig. S5 shows that pores form in DOPC with a much higher probability than in DPhPC membranes. We thus prepared GUVs composed of DOPC:DOPA 1:1 lipids and mapped the membrane potential changes as a function of time (see Supplementary Section S5 for details). From Fig. S6 it can be seen that the SH intensity / membrane potential gradually vanishes after the GUV is exposed to $Cu^{2+}$, $Ca^{2+}$, or $Ba^{2+}$ ions, which means that the asymmetry of the membrane hydration has been restored, with ion transport being the only possible source to achieve this. Note that compared to the FLM of Fig. 1(iii) the ionic strength is much higher in this experiment, which means that some fluctuations are below the detection limit of our instrument.

**Rate of ion translocation correlates with transient membrane potential.**

Using this approach, cation translocation rates are determined (pixel-wise or averaged over the full GUV). Fig. 2A shows the translocation rates for $Cu^{2+}$, $Ca^{2+}$, $Ba^{2+}$ and $Mg^{2+}$ ions. Fig. S7 shows the spread in values obtained within a single GUV for $Cu^{2+}$ and $Ca^{2+}$, which means that also under conditions of ion permeation, the local structure of the membrane and evolution in time and space is important (see Supplementary Section S6). The permeability strongly depends on the ionic species: $Cu^{2+}$ ions transfer the fastest, while $Mg^{2+}$ ions are hardly permeated through the membrane. To test whether the trend in translocation rates ($Cu^{2+}$ > $Ca^{2+}$ > $Ba^{2+}$ > $Mg^{2+}$) is correlated with transmembrane potentials, we reverted to using lipids with phytanoyl tails as in Fig. 1, which render the hydrophobic core highly impermeable (Fig. S5). Fig. 2B shows the obtained spread in membrane potential values for the different divalent cations interacting with the GUVs. Each data point represents the average membrane potential value (colored dot) together with the domain wise spread in membrane potential (colored line). Each symbol represents the data from a single GUV. The data shows that there is again an ion-dependent variation in membrane potential magnitude, with $Cu^{2+}$ leading to the largest transient membrane potential changes and $Mg^{2+}$ leading to the smallest.

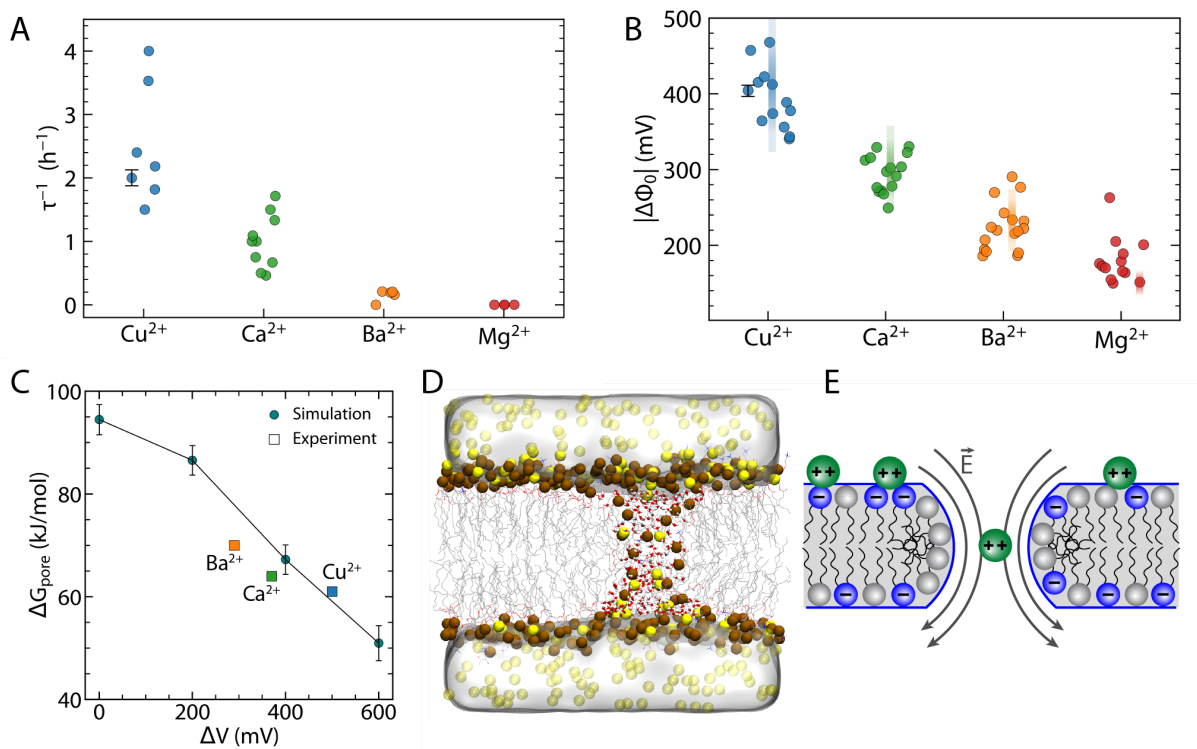

**Figure 2: Divalent cation translocation mechanism.** (A) Translocation rate of divalent cations through 1:1 DOPC:DOPA membrane extracted from intensity decay in SH images. (B) Transmembrane potential values induced by a 5 mM divalent cation solution on one side of 1:1 symmetric DPhPC:DPhPA GUVs. Each dot represents the average potential value of a single GUV and the colored lines represent the domain wise spread in membrane potential. (C) Computed free energy of pore formation in POPC lipid membrane (black curve with dots) together with experimental value for $Ba^{2+}$, $Ca^{2+}$, and $Cu^{2+}$ ions (green square) extracted from measured translocation rate. (D) MD snapshot of an open pore at a transmembrane potential magnitude of 600 mV in a POPC:POPS membrane. Lipid phosphorus atoms are rendered as brown spheres, lipid tails as gray lines and $Ca^{2+}$ ions as yellow spheres. Water molecules inside the membrane are represented as sticks and bulk water as a transparent surface. (E) Schematic of the voltage-induced ion transport mechanism. Transmembrane potential induced by divalent cation binding opens a transient pore through which divalent cations can translocate from one side of the membrane to the other.

Simulated $Ca^{2+}$ ion translocation rates increase with increasing transmembrane potential values for POPC and POPC:POPS membranes, in agreement with the experiment. Fig. S8B shows the permeability times of a single $Ca^{2+}$ ion as computed from the MD simulations and translated to a GUV of a 10 μm radius (see Supplementary Section S7 for permeability calculations). For a transmembrane potential magnitude ranging from 200 mV to 600 mV the rate of translocation increases drastically. This suggests that membrane potential fluctuations have a dominant impact on ion transport and that this transport mainly happens in regions where membrane potential fluctuations reach maximum values. In the case of $Ca^{2+}$ ions, those membrane fluctuations reach an absolute value of ~370 mV. At that transmembrane potential difference, the simulated free energy required to open a pore is on the order of ~70 kJ/mol (Fig. 2C, black curve). In the case of $Mg^{2+}$ ions, the induced transmembrane potential results in a significantly higher pore formation energy, which leads to extremely slow permeation (see Supplementary Section S8). By combining the measured translocation time of $Ba^{2+}$, $Ca^{2+}$, and $Cu^{2+}$ ions (Fig. 2A) with the flux of ions through an open pore obtained from simulations (see Table S1), we calculate the free-energy of pore formation to be ~ 70, 64, and 61 kJ/mol respectively (see Supplementary Section S7 for more details on free energy calculations). Fig. 2C shows that there is good agreement between the experimental and simulated values for the free energy of pore formation. Thus, ion translocation of divalent ions is occurring via the opening and closing of transient membrane pores, which happens due to electrostatic free energy fluctuations induced by intrinsic structural heterogeneities in the hydrated headgroup regions of the lipid membrane. The fact that pores are widely formed by the application of external fields for membrane electroporation gives additional support for the plausibility of our model. Fig. 2D shows the snapshot from MD simulations of an open pore through which $Ca^{2+}$ ions (yellow spheres) translocate from one side of the membrane to the other. Fig. 2E shows the schematic concept of the proposed voltage-induced ion transport mechanism. Our model is distinctly different from the four already existing models, which do not agree with the data, as worked out in more detail in Supplementary Section S1, Figs. S1-S3.

**Conclusion and Summary**

In summary, wide-field high-throughput SH microscopy of lipid membrane hydration together with molecular dynamics simulations were used to study the mechanism of divalent metal cation ($Mg^{2+}$, $Ba^{2+}$, $Ca^{2+}$, and $Cu^{2+}$) permeation through symmetric charged GUVs. GUVs were formed from saturated branched DPhPC:DPhPA 1:1 or unsaturated DOPC:DOPA 1:1 lipids and were brought into contact with solutions of different divalent salts. With SH microscopy we observed membrane potential fluctuations on the surface of GUVs as well as planar free-standing lipid membranes of the same composition. Molecular dynamics simulations show that such variations in membrane potential reduce the energy barrier of transient pore formation and can lead to a passive transport of divalent cations. Other transport models such as the ion-induced defect model or direct ion permeation are not compatible with our data or with previous experiments (see Supplementary Section S1). Using SH microscopy we extracted the average translocation rates for the studied cations and showed that these rates correlate with the average transient membrane potentials induced by interaction of those cations with the lipid membrane. Our findings agree well with the values found from MD simulations. Therefore, we propose ion transport via electrostatic field induced pores as a viable mechanism for passive ion transport.

**Acknowledgments**

**Funding:** This work was supported by the Julia Jacobi Foundation, the Swiss National Science Foundation (Grant 200021-182606-1), the European Union's Horizon 2020 research and innovation program under Marie Skłodowska-Curie grant agreement 721766 (H2020-MSCA-ITN), and European Research Council grant agreement No 951324 (H2020, R2-tension). C.S.P. and J.S.H. were supported by the Deutsche Forschungsgemeinschaft (grant no SFB 1027/B7). **Author contributions:** D.R. and M.E. performed experimental measurements and reproducibility validations. C.S.P. and J.S.H. performed analytical computations and simulations. S.R., J.S.H., D.R., M.E., and C.S.P. wrote the manuscript. S.R. and J.H. supervised the work. Both D.R. and M.E. contributed equally and have the right to list their name first in their CV. All authors contributed to the article and approved the submitted version. **Competing interests:** The authors declare that they have no competing interests. **Data and materials availability:** All data needed to evaluate the conclusions in the paper are present in the paper and/or the Supplementary Materials. Additional data related to this paper may be requested from the authors. The modified GROMACS software that implements the chain reaction coordinate is available at https://gitlab.com/cbjh/gromacs-chain-coordinate.